\documentclass[prl,a4paper,twocolumn,showpacs,floatfix]{revtex4-1}

\usepackage{longtable}
\usepackage{morefloats}
\usepackage[dvips]{graphicx}
\usepackage{color}
\usepackage{epsfig,graphicx,amsfonts,amsbsy}
\usepackage{amsmath,amsfonts,amsthm,amssymb}
\usepackage{appendix}
\usepackage{makeidx}
\usepackage{url}
\usepackage{verbatim}
\usepackage[bookmarksnumbered,pdfpagelabels=true,plainpages=false,colorlinks=true,linkcolor=blue,citecolor=red,urlcolor=blue]{hyperref}
\usepackage[rightcaption]{sidecap}
\usepackage{array}
\usepackage{booktabs}
\usepackage{multirow}
\usepackage{tabularx}
\usepackage{cancel,soul,ulem}

\DeclareMathAlphabet{\pazocal}{OMS}{zplm}{m}{n}

\newcommand{\beps}{\boldsymbol\epsilon}

\begin{document}
\title{Antiferromagnetic Magnonic Crystals}
\author{Roberto E. Troncoso$^{1}$\email{r.troncoso.c@gmail.com}, Camilo Ulloa$^2$, Felipe Pesce$^2$ and A. S. Nunez$^2$}%
\affiliation{ $^1$Departamento de F\'isica, Universidad T\'ecnica Federico Santa Mar\'ia, Avenida Espa\~na 1680, Valpara\'iso, Chile}
\affiliation{$^2$Departamento de F\'isica, Facultad de Ciencias F\'isicas y 
Matem\'aticas, Universidad de Chile, Casilla 487-3, Santiago, Chile}

\date{\today}

\begin{abstract}
We describe the features of magnonic crystals based upon antiferromagnetic elements. Our main results are that with a periodic modulation of either magnetic fields or system characteristics, such as the anisotropy, it is possible to tailor the spin wave spectra of antiferromagnetic systems into a band-like organization that displays a segregation of allowed and forbidden bands. The main features of the band structure, such as bandwidths and bandgaps, can be readily manipulated. Our results provide a natural link between two steadily growing fields of spintronics: antiferromagnetic spintronics and magnonics. 
\end{abstract}

\maketitle

Small deviations in the local magnetization of a magnetic system can propagate coherently in the form of spin waves (SWs) whose associated quantum fields are known as magnons. While neutral regarding its electric behavior they carry a unit quantum of spin, $\hbar$.  Due to the lack of Joule heating associated with its transport they stand out as promising candidates for its application in the context of information processing\cite{Magnon}.  The field of solid state physics concerning the  manipulation, detection and dynamics of the SWs in a magnetic system has been dubbed magnonics\cite{Kruglyak}. The field of magnonics has grown into a well established realm of magnetism and opened new paths in the understanding of magnetization dynamics of complex structures. 
Most of the research in magnetic has focused on spin waves propagating across systems with an overall ferromagnetic order. For example, one of the most studied systems is Ytrium-Iron-Garnet(YIG)\cite{Kruglyak} that, being ferrimagnetic, displays a net magnetic moment on each unit cell. A central theme in the field of magnonics is the implementation of magnonic crystals (MCs), the spin wave analog of photonic and plasmonic crystals, structures with magnetic properties spatially modulated in a periodic fashion. 
In a MC the magnon spectra is organized in the form of bands with associated bandgaps that can be tailored by proper adjustment of the MC properties. A variety of magnonic devices have been proposed that profit from the spin waves as information carriers\cite{Ma,gallardo}, signal filters, phase shifters, isolators, and signal processing elements\cite{Kruglyak}.


In this work we propose to use antiferromagnets (AF) as the basic background material in magnonics devices\cite{MacDonald}. To highlight the potential in the use of antiferromagnetic materials we illustrate two examples of magnonic crystals that can be implemented using antiferromagnetic elements. Among the main results we highlight the possibility of tailoring the magnonic bands by the use of modulations in the magnetic field, the anisotropy of the elements or simply by manipulating the geometry of the system. Our results point towards an effective engineering of the magnonic bands.

Antiferromagnetic based spintronics is a rapidly developing new field by its promising and unique properties for future spintronic devices in magnetism. Despite its lack of macroscopic magnetization, antiferromagnets, interact with spin-polarized currents and can give rise to spintronic effects such as magnetoresistance and spin transfer torques\cite{Nunez,Gomonay,Urazhdin,Hals}, the piezo-spintronic effect\cite{Nunez2} and skyrmion textures\cite{AFMSkyrmion}. There are a number of advantages that AF systems present over ferromagnetic ones regarding potential applications. The lack of stray fields, rapid frequency switching in the terahertz range and its diverse functionalities to be integrated with ferromagnets, are among of the best qualities of antiferromagnets\cite{MacDonald,Gomonay}. The first one relates to the already mentioned lack of net magnetization. Due to this, they do not create magnetic fields which renders local all the interactions involved in its manipulation. Second, the typical time-scale associated with changes in the magnetic structure is several orders of magnitude shorter than the associated with ferromagnetic systems \cite{Fiebig}. This opens a possibility to implement high-speed effects operating in the terahertz range. Finally,  antiferromagnetism is observed more often and at much softer conditions than ferromagnetism, being found even in semiconductors at room temperature \cite{Maca}, this allows us to envision hybrid devices that display features of both electronic and spintronic characteristics. Among the large assortment of antiferromagnetic materials, we have in mind those antiferromagnetic insulators with uniaxial anisotropy such as, for example, NiO\cite{Weber}, MnF$_2$\cite{Kotthaus,Ross}, FeF$_2$\cite{Ohlmann}. Spin waves in AF have been studied since the dawn of quantum mechanics. Both from the theoretical point of view\cite{Anderson} as from the experimental\cite{Shull}. There is a great deal knowledge accumulated over the decades involving the spectra of spin waves in a variety of AF. This opens a window of opportunity for effective control of the magnon degrees of freedom.

We start off our discussion by stating the basic features of our model. We study the dynamics of the staggered magnetization field in a spatially modulated antiferromagnet\cite{Hals,Haldane,Lifshitz}. In terms of the microscopic exchange energy, $J$, lattice constant $\ell$, coordination number $z$ and the uniaxial anisotropy $D$, the free energy density, $F$, for this system can be expressed\cite{Hals} as a functional of the staggered magnetization and magnetization fields, $\bf n$ and $\bf m$ respectively:
$$
F=\left[\frac{a}{2}{\bf m}^2+\frac{A}{2}\sum_i(\partial_i {\bf n})^2-\frac{K_z}{2}\left({\bf n}\cdot\hat{z}\right)^2-{\bf H}\cdot{\bf m}\right],
$$ where $a=4zJS^2/\ell^3$, the homogeneous exchange energy; $A=zJS^2/2\ell$, the exchange stiffness; $K_z=2DS^2/\ell^3$ the anisotropy (easy axis); and ${\bf H}=g\mu_B {\bf B}/\ell^3$ the external magnetic field. These parameters have spatial dependencies whose specific form will be specified later on.
In $F$ the fields are further constrained to obey $\bf n\cdot\bf m=0$ and ${\bf n}^2=1$ at every instant and everywhere within the system.   
The equations of motion can be obtained from a variation of the action where the constraints are enforced with the aid of suitable Lagrange's multipliers. The resulting dynamics is ruled by:
\begin{eqnarray}
\dot{{\bf n}}&=&\gamma{\bf f}_m\times{\bf n},\label{eq: Eqs. Motion}\\
\dot{\bf m}&=&\gamma\left({\bf f}_n\times{\bf n}+{\bf f}_m\times{\bf m}\right)\label{eq: Eqs. Motion2},
\end{eqnarray}
where $
{\bf f}_m=-a{\bf m}+{\bf n}\times\left({\bf H}\times{\bf n}\right)$, $
{\bf f}_n=A{\bf n}\times(\nabla^2{\bf n}\times{\bf n})+K_z({\bf n}\cdot\hat{z}){\bf n}\times(\hat{ z}\times{\bf n})-\left({\bf n}\cdot{\bf H}\right){\bf m}$, and $\gamma$ is the effective gyromagnetic ratio. 

To determine the spin waves we will consider small variations around the Neel vector and the canting field. Since the canting field ${\bf m}$ is itself small respect to the local magnetic moment we linearized with respect to it. Considering ${\bf n}={\bf n}_0+\delta{\bf n}({\bf x},t)$ and ${\bf m}={\bf m}({\bf x},t)$ we expand up to first order the equations of motion (Eq. (\ref{eq: Eqs. Motion}) and (\ref{eq: Eqs. Motion2})) as
\begin{align}
\frac{\delta\dot{{\bf n}}}{\gamma}&=-a{\bf m}\times{\bf n}_0-\left({\bf n}_0\cdot{\bf H}\right)\delta{\bf n}\times{\bf n}_0\label{eq: eqMotionSW}\\
\frac{\dot{\bf m}}{\gamma}&\nonumber=A\nabla^2\delta{\bf n}\times{\bf n}_0-K_z({\bf n}_0\cdot\hat{z})^2\delta{\bf n}\times{\bf n}_0\\
&\qquad\qquad\qquad\qquad\qquad-\left({\bf n}_0\cdot{\bf H}\right){\bf m}\times{\bf n}_0\label{eq: eqMotionSW2}
\end{align}

To solve these equations let us look for monochromatic waves in the form:
\begin{align}
\delta{\bf n}({\bf x},t)&=\left({\beps}_1n_1+\beps_2 n_2\right)e^{i{\bf k}\cdot{\bf x}-i\omega t}\label{eq:sw_monochromatic}\\
{\bf m}({\bf x},t)&=\left(\beps_1m_1+\beps_2m_2\right)e^{i{\bf k}\cdot{\bf x}-i\omega t}\label{eq:sw_monochromatic2}
\end{align}
in this representation we have used the constraints to express both fluctuating fields in the plane perpendicular to $\bf n_0$ that is spanned by the mutually orthogonal  (but otherwise arbitrary) vectors $\beps_1$ and $\beps_2$. In this expression $n_{1,2}$ and $m_{1,2}$ are complex coefficients. 
Starting from Eqs. (\ref{eq: eqMotionSW}-\ref{eq:sw_monochromatic2}) in a direct way we can assess the problem of spin waves in a homogeneous antiferromagnet. The results are, naturally, consistent with the well-known dispersion relation\cite{Lifshitz}:
\begin{align}
(\omega\pm \gamma H)^2=a\gamma^2(Ak^2+K_z).
\end{align}
There are two independent solutions. One has a phase difference between $n_1$ and $n_2$ equal to $\pi/2$, while the other has a phase difference equal to $-\pi/2$. These solutions correspond, therefore, to spin waves circularly polarized to the left and to the right. In the absence of magnetic field both branches are degenerated. This degeneracy is split by the magnetic field that shifts the dispersion relation of the right-polarized waves upward and the dispersion relation of the left-polarized waves downward by an equal amount $\gamma H$.
\begin{figure}[htbp] 
   \centering
   \includegraphics[width=3.5in]{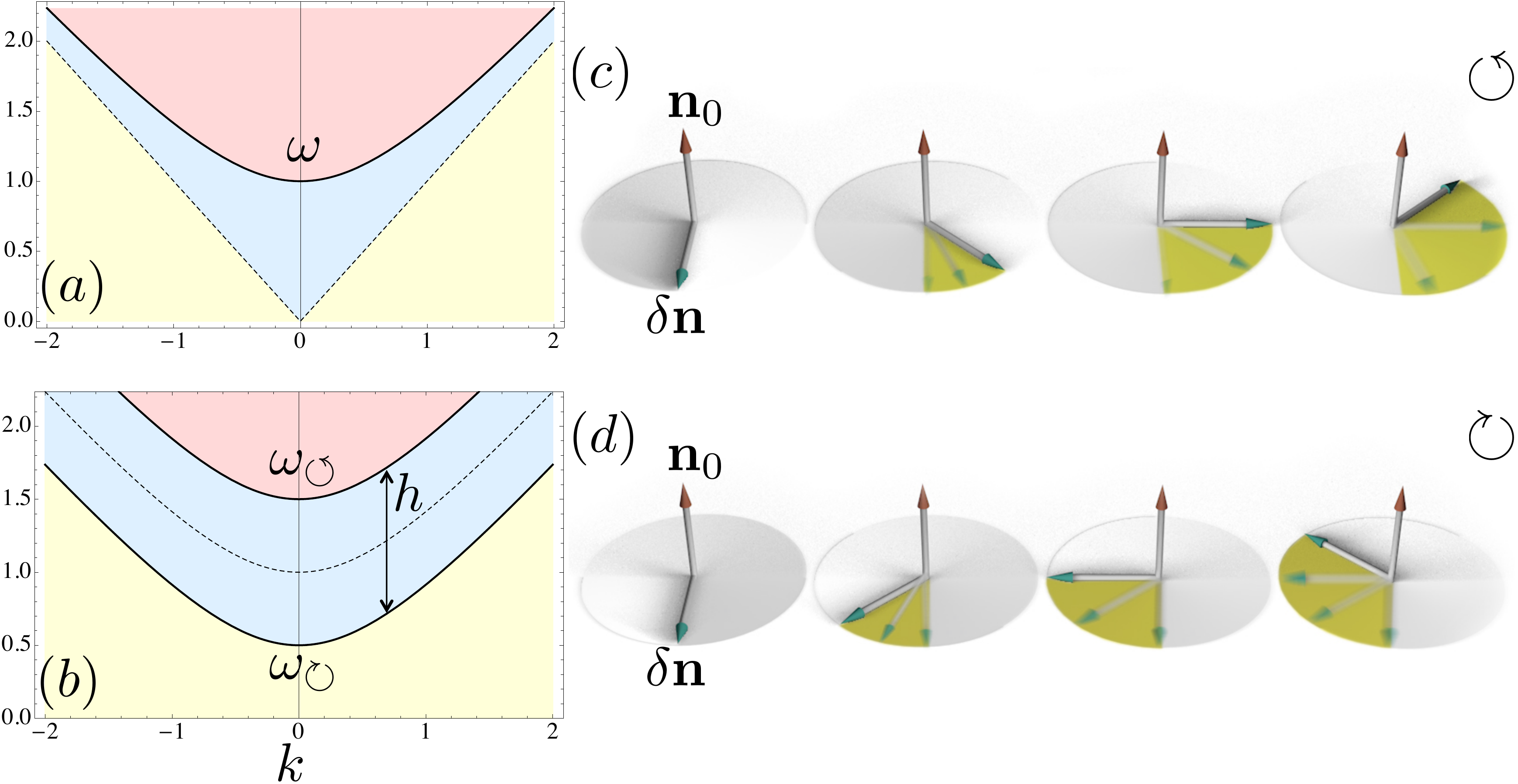}  
   \caption{(a) Magnon dispersion relations for the homogenous antiferromagnet. Without anisotropy the relation is dispersionless (dashed line). Addition of anisotropy raises a gap and changes the dispersion relation to the Klein-Gordon form (full line). In both cases the dispersion relation is doubly degenerated reflecting the two possible polarizations of the spin wave. (b) Addition of a homogeneous magnetic field splits the degeneracy and creates different dispersion relations ($\omega_\circlearrowleft$ and $\omega_\circlearrowright$) for the two opposite polarizations. (c) and (d) Illustration of the two polarizations for the spin wave. The disturbance is perpendicular to the equilibrium Neel vector (${\bf n}_0$) and precesses in a clockwise or anti-clockwise sense.}
   \label{fig:example}
\end{figure}
An important aspect of the spin wave spectra that is encoded by Eqs. (\ref{eq: eqMotionSW}) and (\ref{eq: eqMotionSW2}) is that oscillations in the Neel order parameter $\delta\bf n$ are linked to oscillations in the magnetization. In fact, a quick look at Eqs. (\ref{eq: eqMotionSW}-) and (\ref{eq: eqMotionSW2}) allows us to write:
$$
{\bf m}=-\frac{1}{a}\left(\frac{1}{\gamma}{\bf n}_0\times \delta \dot{{\bf n}}+({\bf n}_0\cdot {\bf H})\delta{\bf n}\right)
$$ 
This simple result is of the great importance since it provides a way to excite and measure antiferromagnetic spin waves by coupling them to the oscillation in the magnetization field that they carry. This coupling to the magnetic degrees of freedom has been used for decades to characterize the spin wave spectra of antiferromagnets\cite{Lvov}.  In this way it is possible to use magnetization probes, such as Faraday's and Kerr's effects or Brillouin light scattering\cite{Kruglyak}, of widespread usage in the field of magnonics, to control and study antiferromanetic spin waves.

Now we are ready to present the main result of the work, the study of different ways in which these antiferromagnetic spin waves can be manipulated with the aid of periodic manipulation of the system parameters. We will see how this manipulation gives rise to magnonic bands that can be tailored with precision. There are basically two essentially different ways to control the antiferromagnetic spin waves. Starting from Eqs. (\ref{eq: Eqs. Motion}) and (\ref{eq: Eqs. Motion2}) we note the possibilities of modulating the system parameters (exchange constants, anisotropy, etc) or the magnetic field. Both parameters can be modulated to generate magnonic crystals and in what follows we will study the peculiarities of each particular modulation. 

Eliminating  ${\bf m}$ from the equations of motion we are led to the following wave equation:
\begin{equation}
\delta\ddot{\bf n}= \nabla^2\delta{\bf n}- \kappa\; \delta{\bf n}+2h\;\delta\dot{\bf n}\times {\bf n}_0+h^2\delta{\bf n},
\end{equation}
where we have set $\tau=1/\sqrt{K_z a\gamma^2}$ as the unit of time  and $\lambda=\sqrt{A/K_z}$, the domain wall width, as the unit of length. In the last equation $\kappa$ and $h$, the dimensionless anisotropy coefficient and $h$ and dimensionless magnetic field ($h=\gamma\tau({\bf n}_0\cdot{\bf H})$) respectively,  are regarded as periodically modulated. For common antiferromagnets\cite{Kotthaus,Ross,Ohlmann,Coey} the value of $\tau$ and $\lambda$ lie in the range of picoseconds  and a few nanometers respectively.

The solutions to the wave equation under a periodic modulation can be expressed in the form of Bloch wave functions $\delta{\bf n}({\bf x},t)={\rm e}^{i{\bf k}\cdot{\bf x} -\omega t}(\beps_1 n_1({\bf x})+\beps_1 n_2({\bf x}))$ where $n_1({\bf x})$ and $n_2({\bf x})$ are periodic functions with the same period as the spatial modulation. The equation of motion, within the Bloch's representation, unfolds into  two coupled equations for  $n_1({\bf x})$ and $n_2({\bf x})$ that can only be fulfilled by choosing a $\pm \pi/2$ phase-shift between them. The waves are, therefore, circularly polarized as in the homogeneous case. Due to the magnetic field there is a splitting between the two circular polarizations. For right polarized waves we have $\delta \dot{\bf n}\times {\bf n}_0=\omega \delta {\bf n}$ while for left polarized waves we have $\delta \dot{\bf n}\times {\bf n}_0=-\omega \delta {\bf n}$. The equation of motion becomes:
\begin{equation}
-\omega^2\delta{\bf n}= \nabla^2\delta{\bf n}-\kappa \delta{\bf n}\pm 2\omega\; h\delta{\bf n}+h^2\;\delta{\bf n}
\end{equation}
where the $\pm$ sign is fixed by the polarization of the  spin wave. This equation, with $\kappa$ and $h$ regarded as periodic functions of space, is the main tool to describe a magnon crystal within an antiferromagnet. Let us give two examples of antiferromagnetic magnonic  crystals that use this equation as the starting point. 

\begin{figure}[htbp] 
   \centering
   \includegraphics[width=3.5in]{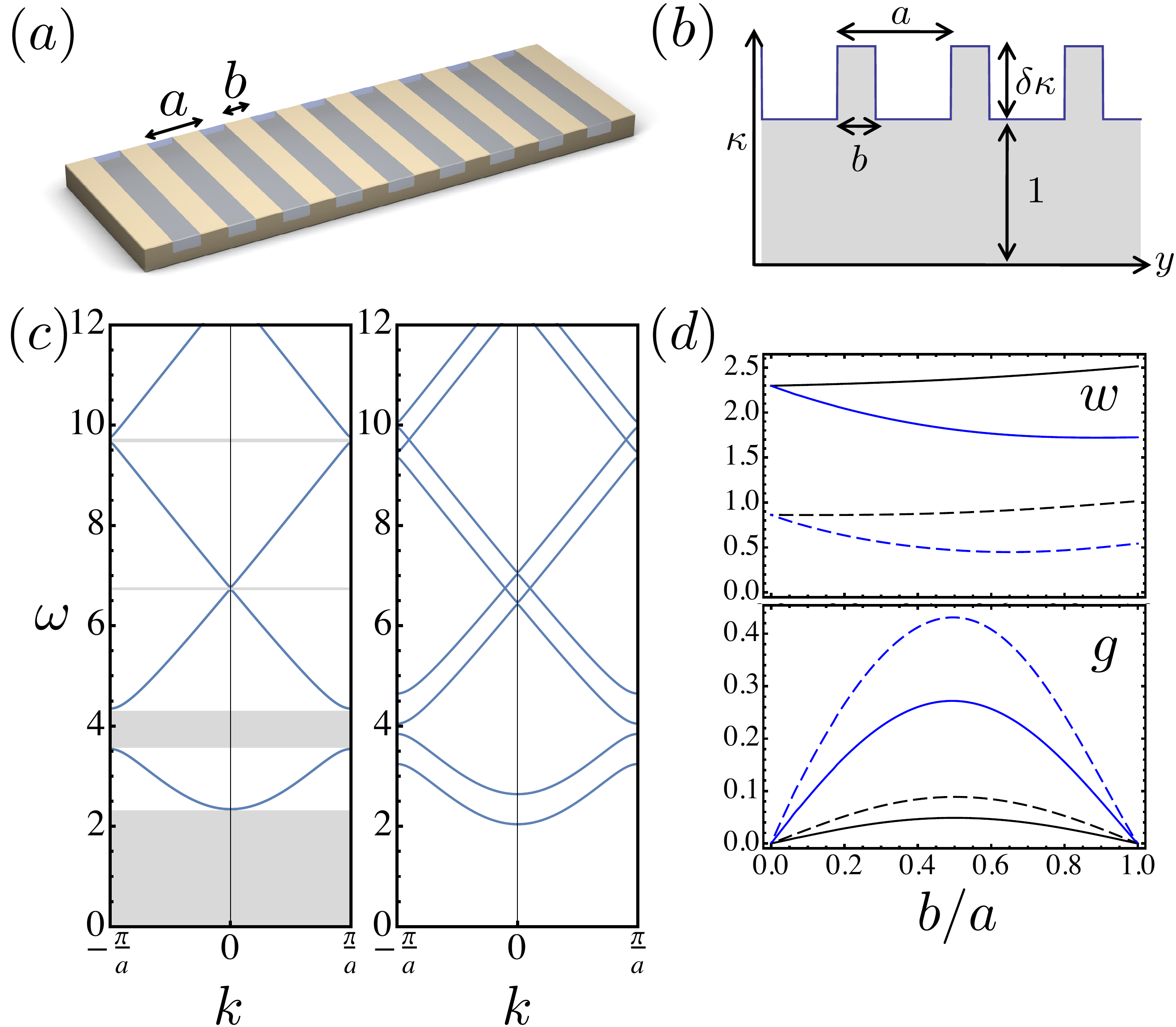} 
   \caption{(a) Model system for a magnonic crystal, a heterostructure with changing anisotropy, illustrating the geometric features.  (b) Simple effective potential that represents the effect of the modulated anisotropy. With the choice of units given in the text, the potential is characterized by a reference anisotropy equal to unity and deviations from it equal to $\delta \kappa$. (c) Left: Magnonic dispersion relation for $a= 1.0$, $b=0.5$ and $\delta\kappa=10$.  Bands are doubly degenerated in account for the different polarizations. Bands of forbidden frequencies are highlighted. Right: Same situation under the action of a uniform magnetic field $h=0.3$. The degeneracy between the different polarization states is broken. (d) Some features of the band structure are displayed as a function of $b/a$. Top: Bandwidth  of the first bands is displayed for different values of the crystal, full lines correspond to $a=1$ and dashed lines to $a=2$. The black and blue lines correspond to $\delta\kappa=-0.5$ and $\delta\kappa=3.0$ respectively. Bottom: With the same parameters we display the bandgap between the first and second bands.}
   \label{fig: anisotropy mediated}
\end{figure}

We begin our discussion by considering the problem in absence of magnetic field.  We will focus on an one-dimensional array with spatially modulated anisotropy as illustrated in Fig. (\ref{fig: anisotropy mediated}). When modulating the spatial anisotropy one can expect that the exchange parameters, $a$ and $A$, should also be modified. Our theory is capable to handle those modulations in a straightforward manner. However, to avoid clumping our discussion with far too many parameters we will focus in a model problem where only the anisotropy is modulated. We have, then,  a series of slabs of width $b$ with different anisotropies are arranged with period $a$. Magnons that propagate along direction transverse to the slabs experience a periodic modulation of the anisotropy parameter thereby giving rise to magnonic bands. This situation can be modeled by the equation 
\begin{equation}
\delta\ddot{\bf n}=\nabla^2\delta{\bf n}- \kappa(y) \delta{\bf n}.
\end{equation}

This equation shows complete degeneracy for the different polarizations. 
Searching for plane waves along the $x$ direction, with wave number $k_x$, we find:
$$
(\omega^2-k^2_x)\delta n_\pm=-\frac{{\rm d}^2}{{\rm d}y^2}\delta n_\pm +\kappa(y)\delta n_\pm.
$$
%
In the physical problem at hand the anisotropy is modulated in a piece-wise constant fashion. The background anisotropy of the system is chosen as the basis for the dimensionless anisotropy $\kappa$. There are slabs of width $b$ distributed uniformly with period $a$. Within these slabs the anisotropy is $1+\delta \kappa$.  
This solution is equivalent to a Schrodinger equation with periodic piece-wise constant  potential and its solutions can be found in textbooks\cite{Kittel}. 
 In Fig. (\ref{fig: anisotropy mediated}) we present the results for a variety of system parameters. A glance at Fig. (\ref{fig: anisotropy mediated}) allows us to state the main results that are (a) double degenerated band structure in account for different polarizations; (b) the appearance of forbidden energies bands; and, (c) the appearance of bands of allowed energies with characteristic bandwidths. As shown in the Fig. (\ref{fig: anisotropy mediated}d) those features can be controlled by an appropriate selection of the parameters of the magnonic crystal. The band structure can be further controlled by exposing the system to the effects of a magnetic field that results in a splitting of the degeneracy of the bands. This is shown in Fig. (\ref{fig: anisotropy mediated}c).

We now consider a magnonic crystal mediated by magnetic field. 
\begin{figure}[htbp] 
   \centering
   \includegraphics[width=3.5in]{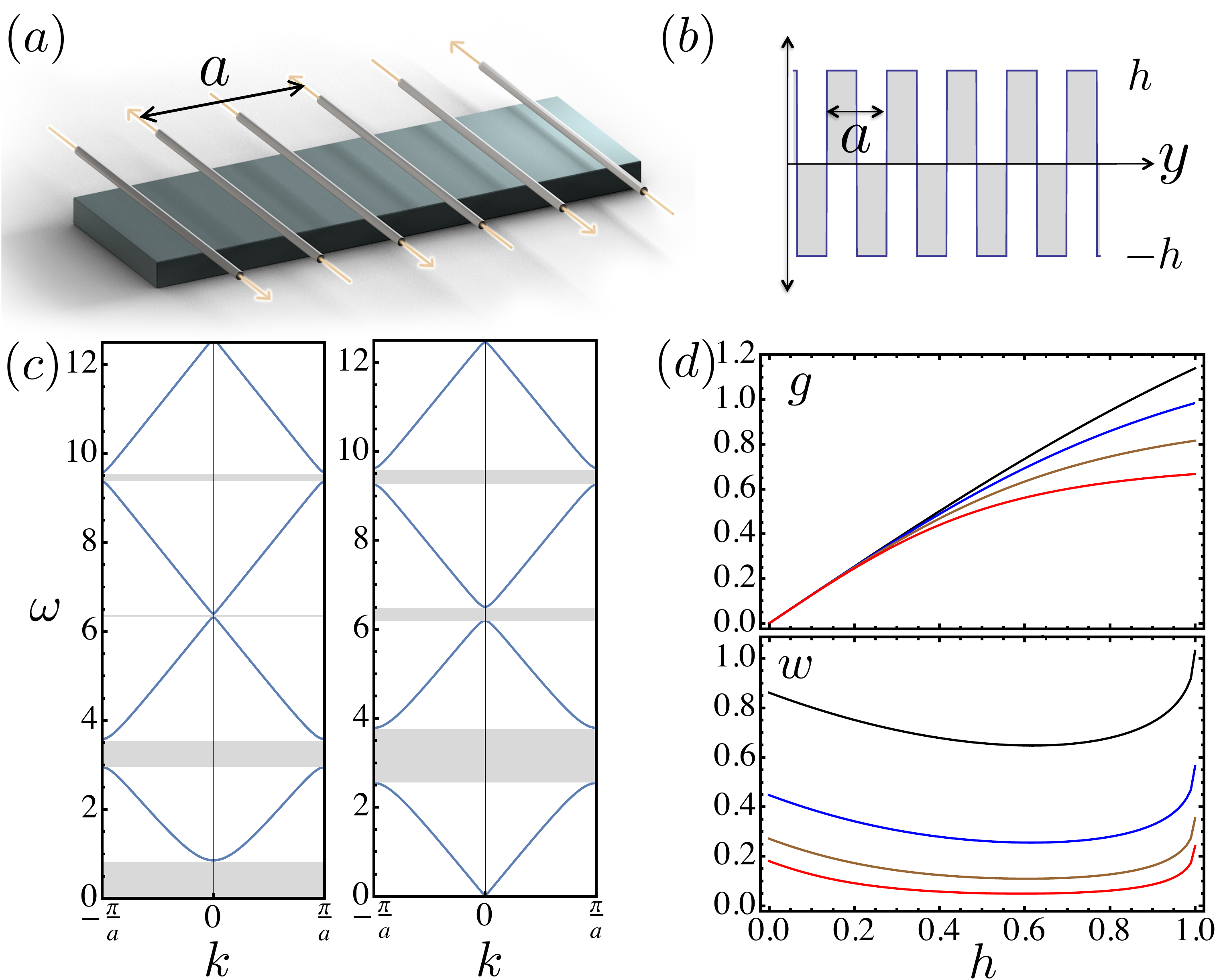} 
   \caption{(a) Arrangement of wires on top of a two-dimensional antiferromagnetic sample. The magnetic field they generate form a magnonic crystal; (b) The system is characterized by spatially modulated magnetic field that oscillates between two extrema $\pm h$ within a period $a$; (c) Left: band structure for $a=1$ , $\kappa=1$ and $h=0.5$. Right:  Band structure for $h=\kappa=1$. The lowest band minimum reaches zero, signaling the spin-flop instability; (d) As a function of the magnetic field strength we display the band structure parameters. Top: The band gap for $a=2$, 3, 4, and 5; Bottom: band width of the first band for the same parameters.}
   \label{fig:field mediated}
\end{figure}
In this system a two dimensional antiferromagnet is exposed to a periodically modulated external magnetic field. While the details of the generation of this magnetic field are irrelevant for the conclusions we are going to draw, we can picture the following arrangement: locate the antiferromagnetic system underneath a periodic array of wires as depicted in Fig. (\ref{fig:field mediated}a). The situation that we propose consists on having a current propagating across the wires. The magnitude of the current across each wire is constant while the direction of the current is changed between consecutive wires. In this way the Oersted field generated by the array of wires acts on the antiferromagnet in the form of a spatially periodic magnetic field that enters into the wave equation (Eqs. (\ref{eq: eqMotionSW}-\ref{eq: eqMotionSW2})). 
To fix ideas on the nature of this equation we approximate the field by a piece-wise constant behavior with values $\pm h$.  
Assuming a plane wave behavior along the $x$-direction we find:
$$
(\omega^2-k^2_x+\kappa+h^2)\delta n_\pm=-\frac{{\rm d}^2}{{\rm d}y^2}\delta n_\pm\;\pm\;2h\omega\delta n_\pm
$$ 
This problem is solved following the standard procedure described in \cite{Kittel}, in the same way as our previous discussion. The main results are displayed in Fig. (\ref{fig:field mediated}). The periodic magnetic field gives rise to a band structure with characteristic bandgaps and bandwidths that are characterized in Fig. (\ref{fig:field mediated}d). As we increase the strength of the magnetic field, the lowest point in the first band decreases continuously from the value at zero field. This trend leads to an instability at $h=\sqrt{\kappa}$ when the lowest band touches the bottom of the axis. For fields greater than this critical value the ground state is distorted in what is known as the spin-flop transition\cite{Ross,Lvov}.

We have discussed the possibility of implementing magnonic crystals in the context of antiferromagnetic spintronics. We proposed two complementary methods to achieve control over the magnonic degrees of freedom: first by controlling the anisotropy properties of the magnon system and second by exposing the antiferromagnet to a periodically modulated magnetic field. In both cases we discussed quantitatively the properties of the resulting magnon spectra and showed how it led to a band-like structure of allowed and forbidden bands whose quantitative features can be tailored by proper adjustment of the parameters of the magnon crystal. This proposal bridges together the  two rapidly developing fields of magnonics and antiferromagnetic spintronic.

The authors acknowledge funding from Proyecto Fondecyt No. 1150072, Center for the Development of Nanoscience and Nanotechnology CEDENNA FB0807, and by Anillo de Ciencia y Tecnonolog\'ia ACT 1117.

\bibliographystyle{elsarticle-num}

\end{document}